\documentclass{revtex4}
\usepackage[latin1]{inputenc}
\usepackage[OT1]{fontenc}
\usepackage{graphics,graphicx}
\def\Journal#1#2#3#4{{#1} {\bf #2}, #3 (#4)}

\def\PRD{{\em Phys. Rev.} D}

\def\PR{{\em Phys. Rev.}}

\def\be{\begin{equation}}
\def\ee{\end{equation}}
\def\bea{\begin{eqnarray}}
\def\eea{\end{eqnarray}}

\begin{document}
\vspace*{4cm}
\title{Observational constraints of a symmetric Milne universe}

\author{A. Benoit-L{\'e}vy }

\affiliation{CEA-Saclay/IRFU \& CSNSM\\
Universit\'e Paris Sud, bat. 104\\
91405 Orsay cedex, France}

\author{G. Chardin}
\affiliation{CSNSM\\
Universit\'e Paris Sud, bat. 104\\
91405 Orsay cedex, France}

\begin{abstract}
The Standard Model of cosmology states a surprising composition of the Universe, in which ordinary matter accounts for less than 5\%. The remaining 95\% are composed of $\sim$70\% Dark Energy and $\sim$25\% Dark Matter. However, those two components have never been identified and remain a challenging problem to modern cosmology. One alternative to the concordance model could be the symmetric Milne universe, composed of matter and antimatter (supposed to have negative mass) in equal quantities. We will present the effects of these hypothesis on classical cosmological tests such as primordial nucleosynthesis, CMB, or Type Ia supernovae and show that this model is in remarkably good agreement with observations.
\end{abstract}

\maketitle

\section{Introduction}

According to the Standard Model of cosmology, our Universe in composed of only 5\% of normal matter (baryons), $\sim$25\% of Dark Matter and  $\sim$70\% of Dark Energy. Although this description is a remarkably good fit to the cosmological observations, it provides no indication on the nature of Dark Energy and requires a high level of fine tuning. Before accepting such a strange universe, it is necessary to study possible simpler alternatives.

In this contribution, we study the non-standard cosmology of a universe with equal amounts of matter and antimatter. In addition, antimatter is supposed to have a negative active gravitational mass. Matter and antimatter are separated, forming an emulsion so that at scales larger than the typical emulsion size, the universe is gravitationally empty and is then adequatly described by the Milne cosmology. We will first describe this symmetric Milne universe and expose our motivations for attributing negative mass to antimatter. Then we will show that this model is in surprisingly good agreement with main cosmological tests.

\section{The symmetric Milne universe}

Motivations for attributing a negative active gravitational mass to antimatter come from the work of B. Carter~\cite{Carter68} on Kerr-Newmann geometry describing charged rotating black holes.
  As noted initially by Carter, the Kerr-Newman geometry with the mass, charge and spin of an electron bears several of the features expected from a real electron. In particular, the Kerr-Newman ``electron" has no horizon, presents automatically the $g=2$ gyromagnetic ratio and has a ring structure with radius equal to half the Compton radius of the electron. Additionally, this geometry presents charge and mass/energy reversal symmetries that strongly evoke the CP and T matter-antimatter symmetry~\cite{Chardin97}: when the interior of the ring is crossed, a second $R^4$ space is found where charge and mass change sign. Therefore, starting from an electron with negative charge and positive mass as measured in the first $R^4$ space, we find in the second space a ``positron" with positive charge and negative mass. 

The Milne cosmology also appears as a natural universe as it has a flat spacetime and hyperbolic open space, to be compared to the standard universe which states a flat space and curved spacetime.

\begin{figure}
\begin{minipage}{.45\linewidth}

\includegraphics[width=\columnwidth]{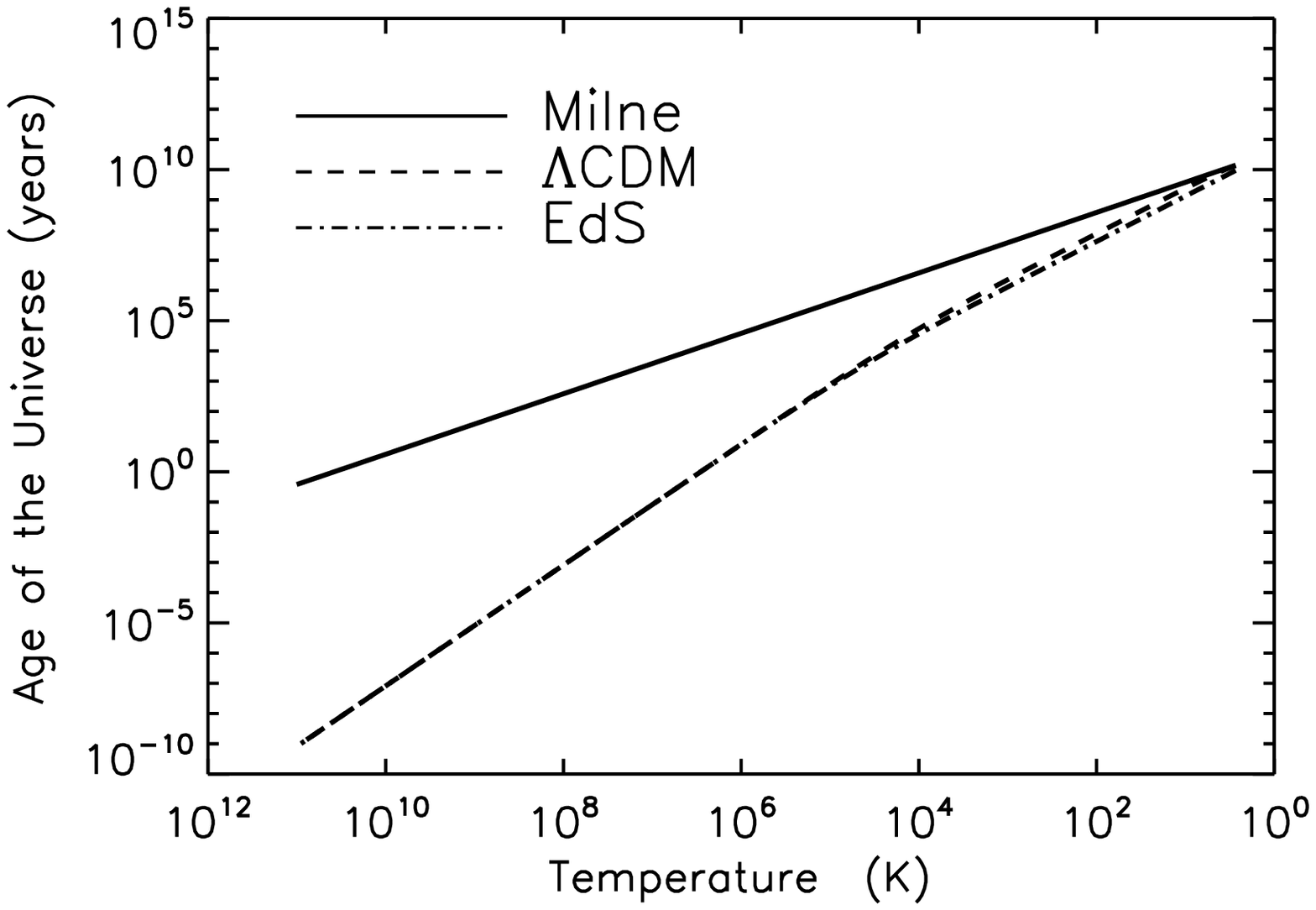}
\caption{\label{age} Age of the Universe vs temperature for different cosmologies.}

\end{minipage}
\begin{minipage}{.45\linewidth} 

\includegraphics[width=\columnwidth]{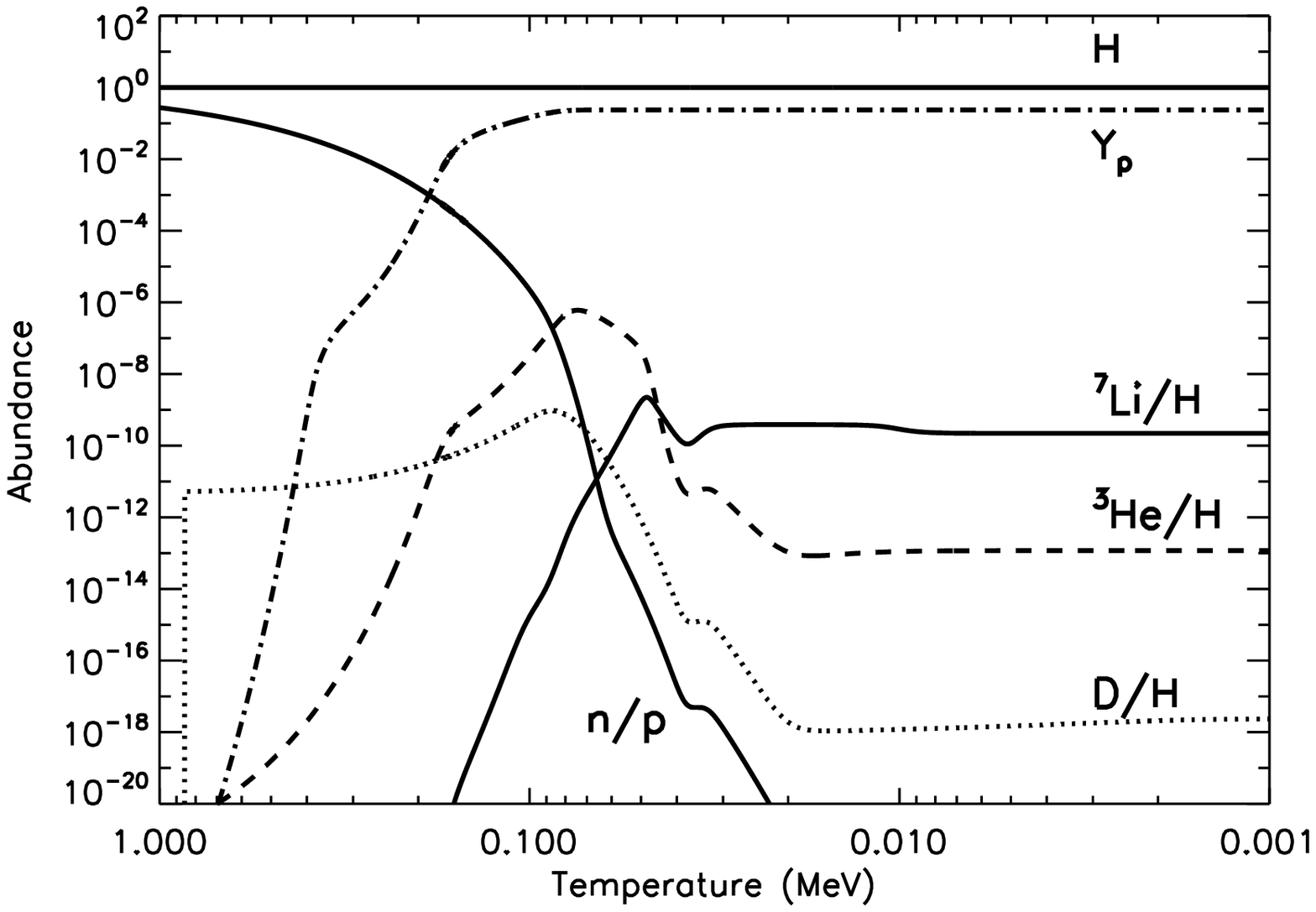}
\caption{\label{mbbn}Abundances of light elements obtained in the Milne Universe with a baryon to photon ratio $\eta=8\times 10^{-9}$.}

\end{minipage}\hfill

\end{figure}

The temperature evolves in a Milne universe as $T \propto t^{-1}$ instead of $T \propto t^{-1/2}$ in the standard model. This leads to very different timescales at high redshift as shown in figure~\ref{age}. For instance, at the epoch of primordial nucleosynthesis, the Milne universe is $10^7$ times older than the standard universe at the same temperature. Also, at recombinaison, the Milne universe is 14 millon year old (instead  of 380,000 years for the $\Lambda$-CDM).

 Today, the age of the Milne universe is simply given by $t_0=1/H_0= 13.9$ billion years with the standard value of $H_0=70\rm{km}/\rm{s}/\rm{Mpc}$, identical to the age of the $\Lambda$-CDM model. This value solves the age problem of the Universe in an elegant way without requiring a Dark Energy component. Another motivation is that there exists no horizon in the Milne cosmology, and that solves, without free parameters, the horizon problem that requires in $\Lambda$-CDM model the introduction of inflation.

In the following, we will study the main cosmological tests of  nucleosynthesis, SNe Ia luminosity distance and  CMB for the symmetric Milne universe.

\section{Primordial nucleosynthesis}

Thermal primordial nucleosynthesis in a linearly evolving universe has been investigated~\cite{Lohiya,Kap} in the general case of power-law cosmologies. These authors have noted that despite the tremendous modification in timescale (three years instead of three minutes), adequate amounts of $^4\rm{He}$ and $^7\rm{Li}$ can be created thermally. Due to this very slow evolution, weak interactions decouple at a temperature around $T \approx 80 \rm{keV}$ instead of the classical value of $T \approx 1 \rm{ MeV}$. When deuterium finally survive its photodisintegration, free neutrons (very few though, in the ratio $n/p=e^{-(m_n-m_p)/T}$) supply the nuclear reactions network and finally end into helium nuclei. These neutrons are regenerated from protons as thermal equilibrium is maintained. Given the huge amount of time available, this mechanism gives the observed abundance for $^4\rm{He}$ at the condition  that the baryon density, characterized by the ratio of the baryon to photon density $\eta$, is increased by an order of magnitude to the value $\eta\approx 8\times 10^{-9}$, to be compared to the standard value of $\eta =  6 \times 10^{-10}$. The theoretical prediction of the abundances are shown in figure~\ref{mbbn}. As noted by Kaplinghat {\it et al.}~\cite{Kap}, $\rm{D}$ and $^3\rm{He}$ are strongly destroyed and need to be somehow produced by another mechanism, which seemed to be hardly possible within the framework of standard cosmology.

However, studies of big-bang nucleosynthesis in the presence of regions of antimatter~\cite{elina,rehm} have shown that diffusion of nucleons from antimatter zone to matter zone (and the inverse process) causes annihilation and photodisintegration at the borders of matter and antimatter domains and thereby provides a means to produce $\rm{D}$, $^3\rm{He}$, and $^6\rm{Li}$ up to the observed abundances. 
More precisely, production by diffusion occurs during a small temperature window, around $T \approx 1\rm{keV}$ when the proton and antiproton diffusion length become of the order of $\approx 10^{-4}$ of the domain size.

We find that, in order to produce deuterium at the level of $\rm{D}/\rm{H} \approx 3 \times 10^{-5}$, the typical size of domains should be of the order of $10^{14} \rm{m}$ at $1 \rm{keV}$. Following Rehm \& Jedamzik~\cite{rehm}, we can also compute the $^6\rm{Li}$ abundance produced by non-thermal reactions, which is correlated to the deuterium abundance.

\section{Type Ia supernovae}

In 1998~\cite{Riess,Perlmutter}, the discovery that distant supernovae were dimmer than expected in an Einstein-de~Sitter universe was interpreted as the proof that the expansion of the Universe was accelerating due to the contribution of a tentative Dark Energy component. We show here that the Milne universe, which neither accelerates nor decelerates is not contradicted by the SNe Ia data. 

\begin{figure}[h]
\begin{minipage}[l]{.45\linewidth}
\includegraphics[width=\columnwidth]{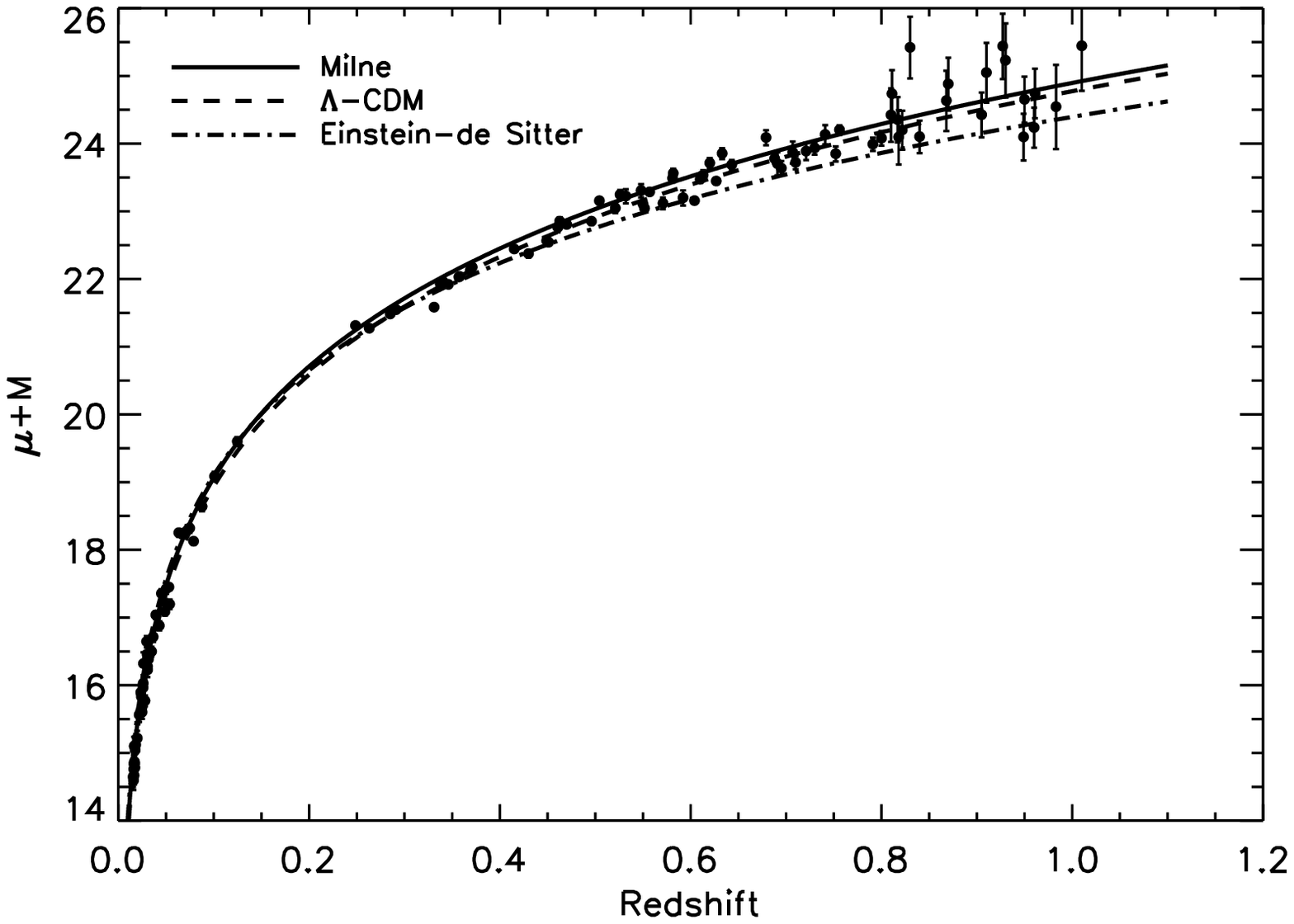}
\end{minipage}
\begin{minipage}[c]{.45\linewidth} 
\includegraphics[width=\columnwidth]{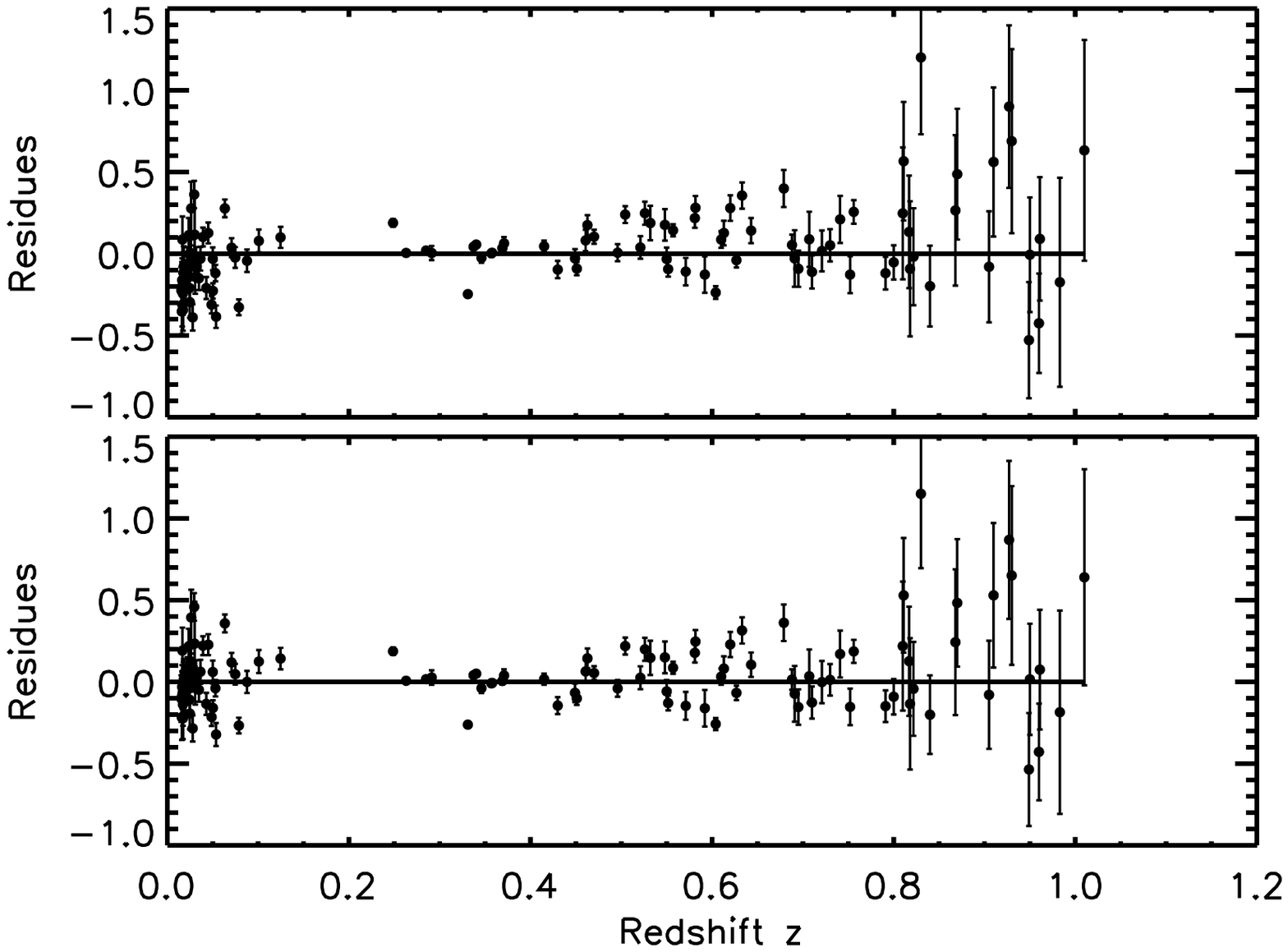}
\end{minipage}\hfill
\caption{\label{hub}\label{res}{\it Left}. Hubble diagram for Milne, $\Lambda$-CDM and Einstein-de Sitter models. While the Einstein-de Sitter universe is clearly excluded, the difference between the Milne and  $\Lambda$-CDM universes is much subtler. {\it Right}. Residuals of the Milne (top) and the $\Lambda$-CDM (bottom) adjustment for the first year SNLS data. It is clear from this figure that the difference between the two models is marginal within the present systematic errors. }
\end{figure}

We used the data of the first year release of the SNLS collaboration~\cite{SNLS} and fitted them to the Milne model. The resulting Hubble diagram is presented in figure \ref{hub}. Whereas the Einstein-de~Sitter model appears excluded, the difference between the Milne and $\Lambda$-CDM models is much subtler. The main difference comes from the value of the absolute magnitude which is fitted as a free parameter alongside the cosmology. In the Milne model, this absolute magnitude is larger by $\approx 0.1$ magnitude. The residuals of the two models are presented in figure~\ref{res}.

Even with the high quality sample of the SNLS, it is very hard to distinguish between these two models and larger statistics and a better comprehension of systematics may be required to finally decide between Milne and $\Lambda$-CDM.

\section{Position of the first acoustic peak of the CMB}

A major test of cosmology today is the Cosmic Microwave Background. Recent CMB experiments tend to show that the space is flat in our Universe and this prediction comes from precise measurements of the position of the first acoutic peak observed at the degree scale.
Surprisingly enough, the Milne universe, which has an open space, also predicts the degree scale for the acoustic peak. More precisely, the angular scale of the first acoustic peak corresponds to the angle under which is seen the sound horizon at recombination:
\begin{equation}
\theta=\frac{r_s(z_*)}{d_A(z_*)},
\label{eq:eqn1}
\end{equation}
$r(z_*)$ being the sound horizon at recombination: $r(z_*)=\int c_s d\eta$, and $d_A$ the angular diameter distance.

 Naively, one would expect a very small value for this angle in the case of the Milne universe due to its open geometry. But in fact, there happens to be a compensation as the Milne universe is 37 million year old at decoupling and because of the integration of the speed of sound with respect to conformal time. Sound waves are generated in the Milne universe by the matter-antimatter annihilation occuring during the period $50 \rm{MeV} - 1\rm{keV}$. Note that annihilation is massive between the QGP transition and $\sim 50 \rm{MeV}$ but only result in heating ofthe medium and not in coherent sound waves.

A more precise calculation yields $\theta_{\rm{{Milne}}}\approx 1.1º$, the degree scale, close to the observed value.

\section{Conclusion}

Motivated by the rather strange and unexplained composition of the Universe and by the symmetries of the Kerr-Newmann solutions in General Relativity, we have studied a natural cosmology with a restored symmetry between matter and antimatter. This model happens to be in good agreement with Big-Bang nucleosynthesis and type Ia supernovae. A preliminary study of the CMB has shown that the first acoustic peak is at the degree scale. All these positive elements are a motivation to study in more details the full CMB spectrum as well as other cosmological tests. Despite unresolved questions, the symmetric Milne universe appears to be a possible alternative to the Standard $\Lambda$-CDM model and its ill-justified Dark Energy and unobserved Dark Matter components.

\section*{Acknowledgments}
It is a pleasure to acknowledge fruitful discussions with N. Fourmanoit, the members of the SNLS collaboration and J.J. Aly. We would like to thank A. Coc, K. Jedamzik and E. Keihanen for the use of their code.

\end{document}